\documentclass[9pt,shortpaper,twoside,web]{ieeecolor}
\usepackage{generic}
\usepackage{cite}
\usepackage{amsmath,amssymb,amsfonts}
\usepackage{algorithmic}
\usepackage{graphicx}
\usepackage{textcomp}
\usepackage{setspace}
\usepackage{tocloft}
\usepackage{caption}
\usepackage{subcaption}
\usepackage{multirow}
\usepackage{adjustbox}
\usepackage[acronym]{glossaries}

\def\BibTeX{{\rm B\kern-.05em{\sc i\kern-.025em b}\kern-.08em
    T\kern-.1667em\lower.7ex\hbox{E}\kern-.125emX}}
\makeglossaries
\newacronym{WSI}{WSI}{Whole Slide Image}
\newacronym{IoU}{IoU}{Intersection over Union}

\begin{document}

\title{Weighted multi-level deep learning analysis and framework for processing breast cancer WSIs}
\author{Peter Bokor, Lukas Hudec, Ondrej Fabian, Wanda Benesova 
\thanks{We gratefully acknowledge the support of Siemens Healthineers Slovakia. The concepts and information presented in this paper are based on research, and they are not commercially available.}
\thanks{Peter Bokor (e-mail: bokorpet98@gmail.com), Lukas Hudec (e-mail: lukas.hudec@stuba.sk; corresponding author) and Wanda Benesova (e-mail: vanda\_benesova@stuba.sk) are with the Slovak University of Technology, Faculty of Informatics and Information Technologies, Institute of Computer Engineering and Applied Informatics, Ilkovicova 2, Bratislava, Slovakia, 84216}
\thanks{Ondrej Fabian (e-mail: fabo@ikem.cz) is with the Institute for Clinical and Experimental Medicine, Clinical and Transplant Pathology Centre, Videnska 1958/9, Prague 4, Czech Republic, 140 21 and the Charles University and Thomayer University Hospital, 3rd Faculty of Medicine, Department of Pathology and Molecular medicine, Videnska 800, Prague 4, Czech Republic, 140 59}
}

\maketitle

\begin{abstract}
Prevention and early diagnosis of breast cancer (BC) is an essential prerequisite for the selection of proper treatment. The substantial pressure due to the increase of demand for faster and more precise diagnostic results drives for automatic solutions. In the past decade, deep learning techniques have demonstrated their power over several domains, and Computer-Aided (CAD) diagnostic became one of them. However, when it comes to the analysis of \glspl{WSI}, most of the existing works compute predictions from levels independently. This is, however, in contrast to the histopathologist expert approach who requires to see a global architecture of tissue structures important in BC classification. 

We present a deep learning-based solution and framework for processing \glspl{WSI} based on a novel approach utilizing the advantages of image levels. We apply the weighing of information extracted from several levels into the final classification of the malignancy. Our results demonstrate the profitability of global information with an increase of accuracy from 72.2\% to 84.8\%.
\end{abstract}

\begin{IEEEkeywords}
Breast Cancer, Whole Slide Images, Voted Classification, Pyramid Approach, Deep Learning, Medical Imaging
\end{IEEEkeywords}

\section{Introduction}

Breast cancer is the most common form of cancer among women. In 2020, there were 2.3 million women diagnosed and 680 thousand died globally of breast cancer. After the last 5 years of diagnosing breast cancer, at the end of 2020, there were 7.8 million women alive and treated with this disease.
Half of the cases, there is no proven significant indicator that could have predicted the risk of breast cancer other than age and gender \cite{whobreastcancer}. Most of the developed countries were able to increase the number of survival by prevention and early detection. An essential part of the diagnostic process is the microscopic analysis of tissue extracted by biopsy and stained by hematoxylin and eosin, the dye that highlights cells from the stroma. A pathologist then states a histopathological diagnosis of breast cancer by recognizing the glandular epithelial structures. Unfortunately, this manual process is time-consuming and difficult due to the variance of stained data, similarity of benign and malignant structures, the vast size of histopathology slides, and other environmental influences. The psychological pressure, demand for fast diagnosis, a large number of postponed examinations, especially in the post-covid era, all lead to an increase in demand for an unbiased histopathology analysis support system. It should be free of environmental influence, capable, reliable, and robust enough to assist pathologists with their analyses.

\begin{figure*}[tbh]
    \includegraphics[width=\textwidth]{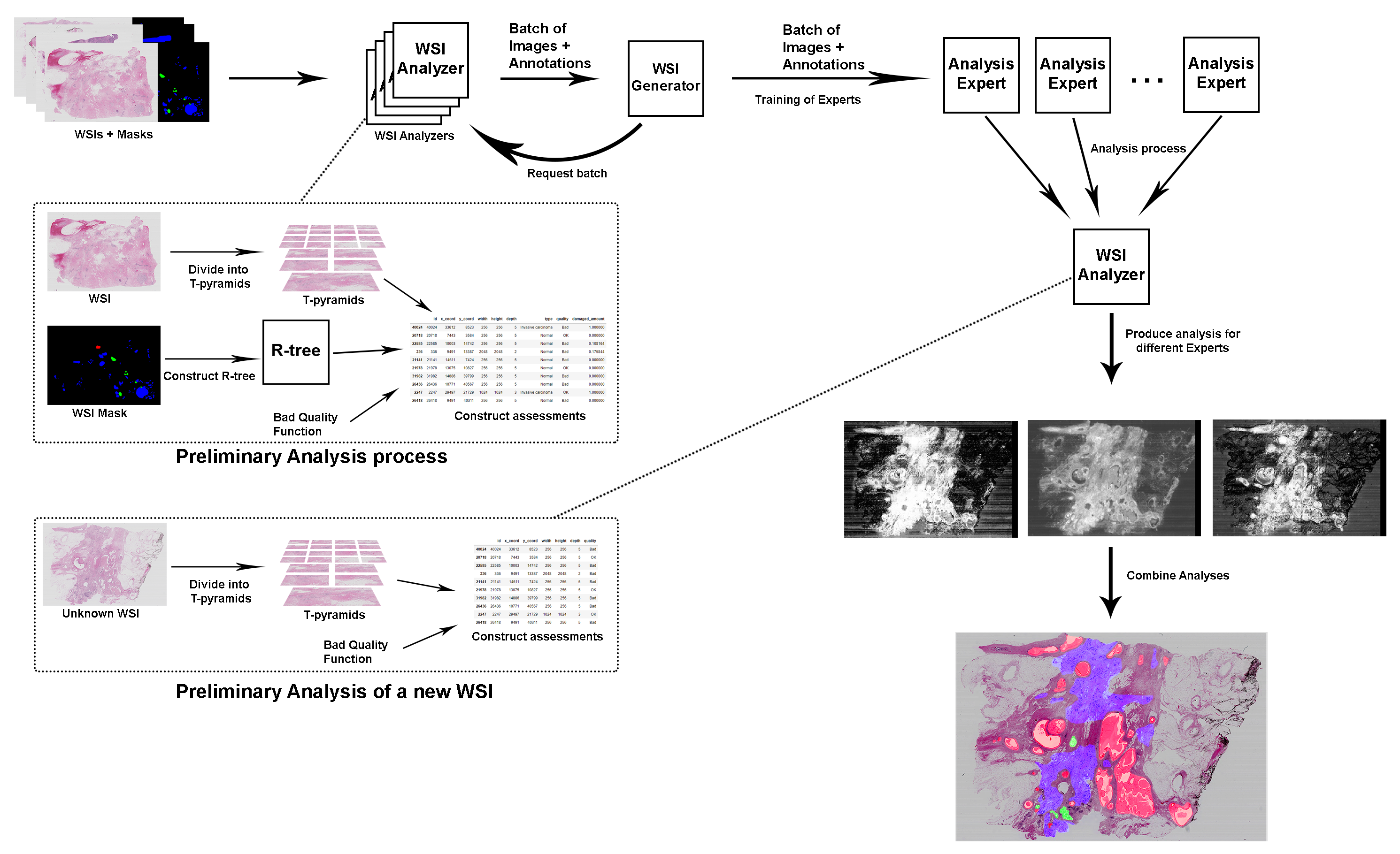}
    \caption{Overview of our \gls*{WSI} processing system - Horizontal flow = training, vertical flow = analysis of new \glspl*{WSI}. The \glspl*{WSI} and their annotations are processed using WSI Analyzer. Annotation regions from external XML files are matched to \gls*{WSI}, bad quality regions are assessed and windows of different magnifications are selected from the \gls*{WSI}. WSI Analyzers are used to provide images with annotations to WSI Generators, which are used to train Experts. Leveraging the flexibility and customizability of the Analyzers and Generators, a general Expert may be trained for numerous amount of tasks. The Experts are then used to produce complex analyses of unknown \glspl*{WSI} by combining simpler partial analyses. Best viewed in color.}
    \label{fig:proposal_flow}
\end{figure*}

We propose an end-to-end 3 steps solution for the automatic processing of Whole Slide Images (WSI). The first step is an effective preprocessing of the \gls*{WSI} images with our proposed framework. The further steps are based on using a sequence of neural network models trained for specific tasks which we call Experts. The last step using one of the mentioned Experts is a novel architecture weighing the influence of different microscopic magnification levels into final classification. An overview of our proposed framework is shown in Figure \ref{fig:proposal_flow}.

We design a module, WSI Analyzer, that processes Whole Slide Images and manages their organization. It matches annotation to the \gls*{WSI}, and hierarchically organizes the data. \glspl*{WSI} processed by WSI Analyzer are subsequently used by another designed module - WSI Generator which produces large sets of data based on multiple parameters.

WSI Analyzer can be also used to produce a completely new analysis, using an Expert. The Expert then applies their knowledge to a new \gls*{WSI} which results in analysis. The produced analysis contains information about the predicted diagnosis that doctors can use as a helping tool or diagnostic support.

We propose several Experts using binary and multi-class classification to generate sufficient and precise diagnostic predictions. We propose an Expert with a novel domain approach for the multi-class classification, that brings attention to different levels of magnification for recognition of different tumor types. All of our experiments are performed with a publicly available dataset BACH ICIAR 2018.

We evaluate our approach quantitatively and empirically. We compare it against state-of-the-art methods in the domain and each other, with a focus on the enhancement. 

Our work and contributions include: 
\begin{enumerate}
    \item Complex analysis of breast tissue \glspl*{WSI} using multiple deep learning models, taking into account contextual and detailed information, simulating analysis approach of a histopathologist.
    \item A deep learning model with a special architecture capable of enhancing multi-class classification prediction results.
    \item An automatic universal framework for processing large datasets and large data that are natural in form of \glspl*{WSI}. The framework is designed to speed-up the extraction of necessary data to train deep learning models to achieve state-of-the-art results. 
\end{enumerate}

\section{Related Work}
Many recent works focus on creating automatic digital pathology systems that support the classification and segmentation of histopathology microscopy data. The approaches could be divided into two streams according to the data used: typical microscopy tissue images; more modern \gls{WSI} slices containing a very high-resolution scan of a magnified tissue sample in a pyramidal structure. The main effort is to build effective automated analysis systems \cite{Litjens2017} with precise predictions. The most promising approaches are based on Deep learning. 

Alongside the ICIAR 2018 dataset, Spanhol et al. \cite{Spanholdataset} create a large dataset called BreakHis consisting of 7909 images containing eight breast tumor types and also provide baseline results using traditional computer vision methods for feature extractions and traditional machine learning methods for classification. Spanhol, in their paper, stresses the complex nature of histopathology data due to the large visual disunity of structures within tumor types, thus making the classification task even more difficult.
Most of the recent literature base their research on the BreakHis dataset, as it contains image slices from 82 patients and four magnification levels.

Bayramoglu et al. \cite{Bayramoglu2016} focus their research on magnification invariant classification of breast histopathology images using a modern approach leveraging the power of deep learning to replace the whole traditional pipeline. Concluding their work, the authors remark that CNNs provide more promising results than handcrafted features in breast cancer classification.

Recently Attalah et al. \cite{Attallah2021} applied five more advanced models for better extraction of texture features, InceptionResnet v2 among them proven as one of the best, with two handcrafted features for breast cancer classification. They also apply an AutoEncoder network to learn optimal representation and reduce the dimensionality of feature vectors created by a fusion of extracted features. They reached a test accuracy of 83.984\% on the ICIAR dataset.

Celik et al. \cite{Celik2020}, on the other hand, experimented with other advanced models to recognize Invasive Ductal Carcinoma in histopathological tissue images. Their DenseNet and ResNet models were pre-trained on ImageNet and finetuned on a dataset with 227542 samples of $50\times 50$ pixels. The best achieved F1 score was close to 94\% and balanced accuracy of 91.57\%.

Other approaches considered the magnification very relevant and tried to modify their approaches to exploit most of the context information provided. 

Gandomkar et al. \cite{Gandomkar2018} proposed a sequential classification framework that finetunes a separate ResNet network, pre-trained on ImageNet, on image patches for each magnification factor from the BreakHis dataset. The first stage is binary categorization to malignant/benign classes and then into eight corresponding subclasses. According to the highest confidence from all predictions on each patient's microscopy samples, their approach predicted the final diagnosis. Even though they downsampled the images from $700\times 460$ to $341\times 224$, they achieved 96.25\% accuracy for eight class categorization. The magnification level accuracy ranged from 82.4\% to 97.96\%.

There were experiments to create specialized neural network architecture with many exciting approaches to increase its attention on context and local information. One such work \cite{Togacar2020} introduces an attention block that employs channel and spatial attention to provide more stable classification results. However, their model's quality and the success rate are questionable due to their training graphs which indicate that their network is overfitted and their validation subset is insufficient.

Han et al.~\cite{Han2017} propose a class structure-based deep convolutional neural network (CSDCNN) capable of overcoming the problems of similar structures in different classes by optimizing the distance of feature vectors of different classes in feature space. They achieve state-of-the-art results both in binary and multi-class classification. Furthermore, they point out that CSDCNN can classify whole slide images by fully preserving global information contained within the images and avoiding the limitations of patch extraction methods.

\glspl{WSI} offer better utilization of context information from microscopy images with their pyramidal multi-scale structure. Recently, Tripathi et al. \cite{Tripathi2021} used annotated \glspl{WSI} and 400 patch samples from the Bach ICIAR dataset. They state that if the prediction needs to be made for the whole tumor or gland, the network cannot lose spatial nor structural information. They transformed the \gls{WSI} into a sequence of patches and extracted features using GoogleNet pre-trained on the ImageNet dataset. This transformation and feature reduction allowed them to apply a bidirectional long-short-term memory (BiLSTM) network architecture. It is questionable if the model suffers the Clever Hans effect due to ImageNet pretraining, but the approach achieved 84.02\% overall accuracy.

More complex analysis using saliency prediction on each magnification level is presented in Gecer et al. \cite{Gecer2018}. Four fully convolutional networks detect saliency maps representing the histopathologists' diagnostic regions of interest. Instead of classifying all patches, the following networks receive only the most salient regions, significantly reducing time complexity. Then the majority voting or trained SVM evaluates downsampled classification and saliency maps to predict the final class for the whole \gls{WSI}.

Another speed-up by removing non-malignant regions proposed  Ni et al. \cite{Ni2019} in their segmentation network using parallel DeepLab network branch classifying pixel-wise malignancy. This patch/pixel-wise approach required a specific hierarchy aware loss function considering the contribution of structures of each segment and image samples with a single label region.

The most similar approach to our proposed method has been published very recently by Feng et al. \cite{Feng2021}. They segment \glspl{WSI} of liver cancer biopsy samples and classify them into final diagnoses applicable to the whole slides. They experimented with several segmentation networks, and the best results achieved U-Net. A similar pyramid approach is applied to multi-scale \gls{WSI} generated by Gaussian pyramid down-sampling. Then seven separate pre-trained U-Nets are finetuned on individual magnification levels. The method cuts the slide into 12 patches of $448\times 448$ and shifts them in 3 directions to eliminate the checkerboard artifacts from combined segmentation by weighting. The reverse pyramid operation projects upsampled segmented \gls{WSI} from all levels into original magnification. The final segmentation map is generated by majority voting of individual thresholded segmentation maps from each level.

According to the discussions with doctors and existing literature, assisting doctors with meaningful analysis would help speed up their workflow. Analyzing only magnified and high-detailed tissue samples is not enough for the pathologist, as these cut-outs often contain little meaningful information about the tissue structures. Furthermore, most existing approaches present pathologists only unexplained classification results produced by a deep learning model. Like doctors, recognizing malignant structures from these small, highly-magnified regions are also difficult for the network, which does not see the context that could provide valuable additional information. The more promising approaches analyze \glspl{WSI} instead of partial cut-outs. However, as stated in Feng et al. \cite{Feng2021} the problematic step in the analysis is to combine the individual segmentation maps from all magnifications. The problems occur in selecting the optimal threshold value for pixel-wise probability and the static number of votes in majority voting.

Our framework focuses on analyzing the \glspl{WSI} and provides multiple contextual information to every magnified window, thus creating a pyramidal analysis by neural network classifying window samples - slide level multi-class segmentation. We introduce an adaptive weighting of magnification levels correcting their contribution to the final combined classification. Pathologists can use the created analysis for further diagnosis, as we recognize four classes in the Bach ICIAR dataset. Furthermore, we aim to employ a method to solve the inter-intraclass variance, leveraging different magnification levels.

\section{Framework for \gls{WSI} Analysis}

Due to its vast size and multiple magnification levels, the \gls{WSI} contains a massive amount of data. Accessing this data, binding annotation to regions, and using it efficiently (generating a batch for training a model must not take too long) is a challenging task. To address this challenge, we propose a two-stage universal framework for complex \gls{WSI} analysis consisting of Data Processing and Data Analysis pipelines. To get the data from \glspl{WSI} slides, we use method calls from a Python version of OpenSlide library.

The data processing procedure consists of two main units - the WSI Analyzer and the WSI Generator. WSI Analyzer is used to manage \gls{WSI} and its annotation and, upon request, produce image patches with associated annotations. These images may subsequently be used by a WSI Generator to train a machine learning model.  

The data analysis pipeline produces analyses for new, previously unseen \glspl{WSI}. As the new slices do not have annotations with diagnoses, WSI Analyzer uses a machine learning model capable of analyzing images to generate new annotations.
WSI Analyser uses the model to infer annotations by classifying pyramidal regions of the WSI with multi-magnifications analysis. WSI Analyser may use multiple models to produce complex combined analysis.

Analyzing the whole \gls{WSI} using multiple magnification levels yields a complex multi-magnifications analysis capturing both high-magnification details and the lower-magnification architecture of tissue which provides a context. Analyzing a \gls{WSI} slice this way is comparable to the analysis performed by a pathologist.

\subsection{Dataset}

The proposed Expert models were trained on the ICIAR 2018 \cite{ARESTA2019122} dataset. The dataset consists of 10 annotated \glspl{WSI} and 20 unannotated. Due to the small amount of data and necessary structures, we utilized nine \glspl{WSI} for the training of our Experts, and the remaining one was used for evaluation. The annotations associated with the \glspl{WSI} contain three classes of damaged tissue - benign, in situ, and invasive carcinomas. Visualization of the annotation may be seen in Figure \ref{fig:mult_res} (left image), where each of the RGB channels represents a single class, the black is normal tissue or background. The annotation may also be used to divide the \gls{WSI} into tumorous and normal regions, producing a binary analysis (shown in Figure \ref{fig:eval_binary}, right). A usual approach to unify the H\&E staining colors is normalization. However, we do not normalize data to experiment with our approach's robustness and spare the computing resources to speedup the analysis process.

\subsection{WSI Analyzer}

The primary function of the WSI Analyzer is providing easy access to multiple magnifications of microscopy scans contained within a \gls{WSI} and registering them to their annotations. WSI Analyzer procedures split the WSI into smaller window blocks using a complete Quadtree (T-pyramid) in order to assess each of the windows' quality meta-information. This structure is stored in memory for future access to the images and contains information about the positions, sizes, types, and information quality (background+fat vs. possibly malignant tissue; we can exclude images that may not contain diagnostically important information) of each window block.

\begin{figure}[htb]
    \centering
    \includegraphics[width=0.8\columnwidth]{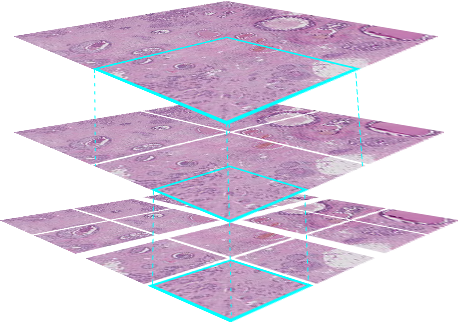}
    \caption{Visualization of first three levels of the T-pyramid. Best viewed in color.}
    \label{fig:qb}
\end{figure}

The window blocks contain patches from all magnification levels (six in ICIAR 2018), creating a pyramidal structure, so we can properly utilize the whole \gls{WSI} and extract as much information as possible. The generation of multi-magnifications T-pyramid structure is visualized in Figure \ref{fig:qb}. Using the T-pyramid structure, we can easily combine specific information - acquired from images with higher magnification with contextual information - from images with lower magnifications.

\subsubsection{Annotation Using the R-Tree}

\begin{figure}[htb]
    \includegraphics[width=\columnwidth]{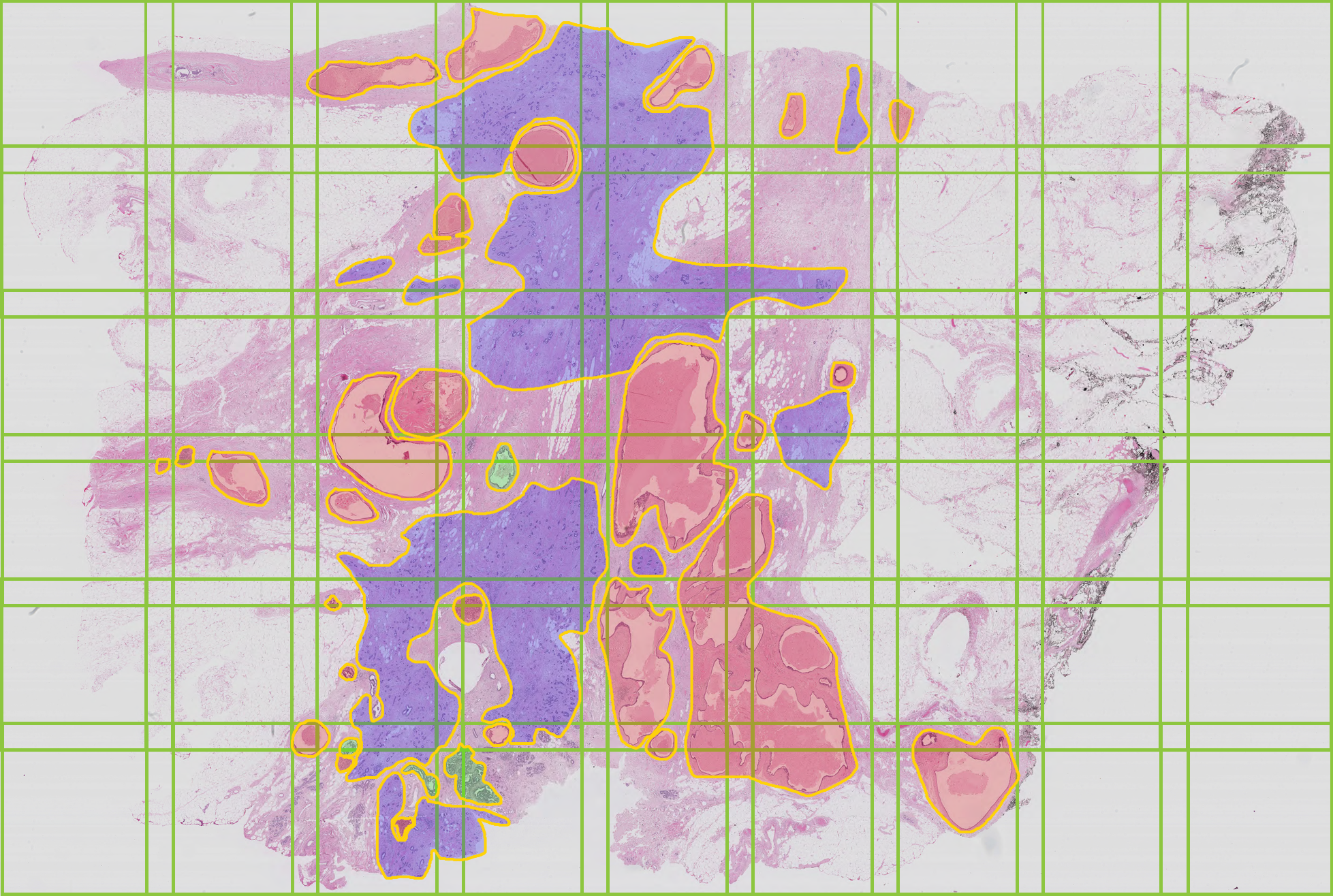}
    \caption{Visualization of first-level square patches intersected with annotation regions. Green windows represent patches, different colors of regions mean different tumor type. Best viewed in color.}
    \label{fig:r_tree}
\end{figure}

With \gls{WSI} and its annotation as an input, the WSI Analyzer constructs an R-tree of the regions contained in the annotation. The R-tree is a search tree over axis-aligned rectangular regions, mainly used for spatial indexing and accessing geographical data, which are not dissimilar to our annotated regions. Using this structure provides a convenient way of performing spatial queries on images in the form of square-shaped regions, we call Quads. Spatially intersecting these square regions with the R-tree index, where annotation information about regions of the \gls{WSI} is stored (as shown in Figure \ref{fig:r_tree}), we achieve a fast way to assess whether an image belongs to at least one tumor region, and if so, how much area of the image belongs to each region. Having access to such data, we can create actual annotations for each image within the T-pyramid by assigning each Quad a class. The resolution of generated patches is arbitrary, and we used the size $256\times 256$ pixels for patches from every magnification level. The smaller patch size of $128\times 128$ contains four times less information than $256\times 256$ and lacks the context, and larger patches increase the demand for computing resources, which is not feasible in hospital integration. Tripathi \cite{Tripathi2021} and several other works used this resolution and noted similar reasons.

\subsubsection{Producing Images and Annotation}

WSI Analyzers' input parameters allow it to select square tissue patches, Quads, along with their assessments to create a batch of images with their corresponding annotations based on multiple parameters: classes to use, magnification levels, information quality, batch size, custom preprocess function applied to each image, size of extracted images. A Parametrized Batch Producer is created to manage the selection of data from each class contained within the \gls{WSI} according to different parameters. The combination of the outputs from these producers then generates a well-balanced mini-batch containing samples from each class. Parametrized Batch Producers may be used with different already mentioned parameters, fine-tuning the batch production to all needs of well-balanced training of our proposed neural network and others trained on \glspl{WSI}.

\subsection{WSI Generator}

We design WSI Generators to produce training data processable by most of the known Deep Learning frameworks and suitable for the training of neural network models. WSI Generators can receive data from multiple WSI Analyzers so that the network model can be trained on a large set of different images. This training strategy is essential for the good generalization of the model. When the Generator object is initialized, all samples from the same classes from all used WSI Analyzers (and all used \glspl{WSI}) are unified and considered into creating the mini-batches. The unification is necessary to ensure that all classes are present for training. The selection and labeling of the classes are also required conditions for consistent one-hot encoding.

The Generator procedures are designed to be applied with different machine learning tasks - multiclass classification, binary classification, and even segmentation, and can extract and generate tissue patches with different labels used by different models.

\section{Deep Learning Experts}

In addition to the data preparation, WSI Analyzer's purpose is to analyze and predict diagnosis annotations for regions of new \glspl{WSI}. The prediction unit of WSI Analyser has to be at least one Expert that is provided during implementation calls while creating the WSI Analyzer. When processing unannotated \gls{WSI}, there is no information to construct R-tree with labels. The labels and quality type are assessed from predictions by analysis experts models. Other processing steps are the same as in previous use-cases.

After the \gls{WSI} analysis process is over, the results are recorded into corresponding T-pyramid Quads. The analysis process creates Quad pyramids of all available magnifications combining details with low magnification architectures, and the Quad structure may resemble an organized 3D graph. Multiple Experts generate multiple analyses assessments, which are contained inside each graph node. The final aggregated analysis depends on the applied Experts and results in a complex analysis of the slide. Analyses may also be generated with every Expert separately.

\subsection{Binary Classification Expert}

The first Expert we design is inspired by the Mask R-CNN's region proposal approach described and demonstrated by Li et al.\cite{Li2019} in a similar domain and produces binary classification of the \gls{WSI}. We apply a simple binary classifier to divide the \gls{WSI}'s T-pyramids into two sets - one that contains diagnostically essential structures of damaged tissue and one that does not and mostly consists of only healthy tissue and fat. 

We train and use this first Expert to quickly and correctly reduce the size of many T-pyramids generated from whole \gls{WSI}. The resulting binary map may theoretically provide enough clues to the doctor while deciding on a diagnosis. Furthermore, it proposes regions with suspected damage for further analysis applying other, more specialized Experts. 

\subsubsection{Binary Classification Expert Design}

\begin{figure}[htb]
    \centering
    \includegraphics[width=0.7\columnwidth]{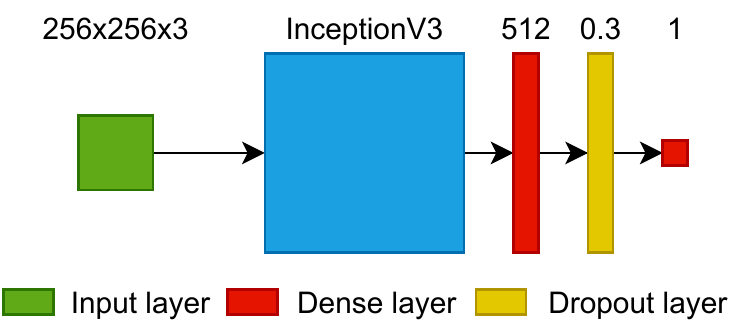}
    \caption{Architecture of the Binary Expert. Best viewed in color.}
    \label{fig:binary_scheme}
\end{figure}

After several experiments with a custom simple convolutional network, we decided to rely on a network with already existing and well-established architecture InceptionV3 \cite{szegedy2016rethinking}, with 1 Dense(512) and 1 Dropout(drop\_rate=0.3) layer added at the end of the network. Scheme of the architecture is shown in Figure \ref{fig:binary_scheme}. Out of multiple experiments using different parameters and configurations, we achieved the best validation accuracy of 0.8662. The optimizer was Adam with the learning rate of $1e^{-6}$, the mini-batch size used was 32, and the model trained for 29 epochs before early stopped. WSI Generator produced 28800 samples for training.

\subsection{Multi-Class Classification Expert}

After filtering out redundant structures, we can focus purely on regions marked as damaged and analyze them further in detail. For this purpose, we design a Multi-Class Classification Expert. The Expert can distinguish between 3 tumor classes contained within the dataset we use - Benign, Carcinoma in Situ, and Invasive Carcinoma.

\subsubsection{Multi-Class Classification Expert Design}

\begin{figure}[htb]
    \centering
    \includegraphics[width=0.7\columnwidth]{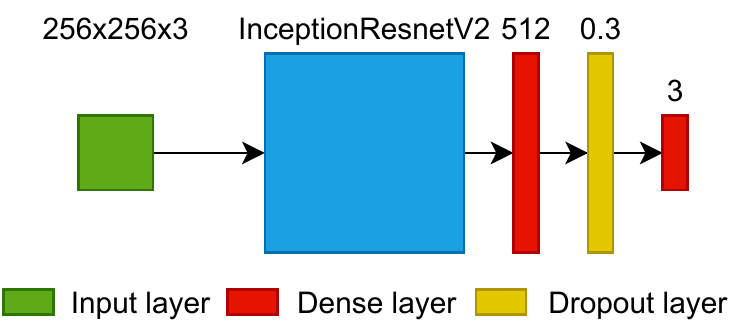}
    \caption{Architecture of the Multi-Class Expert. Best viewed in color.}
    \label{fig:mult_scheme}
\end{figure}

Considering the visual structural problems present within breast tissue microscopy, mentioned by both Spanhol \cite{Spanholdataset} and Han \cite{Han2017}, and also other authors, we utilize the strength of a robust architecture - InceptionResnetV2 \cite{szegedy2017inception}. This architecture provides enough capacity to observe and recognize the differences and similarities between and within the classes to perform satisfactory multi-class classification. Following the same pattern as in the previous experiment, we added 1 Dense(512) and 1 Dropout(drop\_rate=0.3) layer at the network's end. The scheme of the architecture is in Figure \ref{fig:mult_scheme}. The optimizer was Adam with the learning rate of $1e^{-7}$; the mini-batch size used was 18; the model was trained for 167 epochs before early-stopped and achieved the best validation accuracy of 0.62. WSI Generator produced 486000 samples for the training.

\subsection{Weighing Predictions Expert}

The main contribution of our paper is the Weighing Prediction Expert.
Different magnifications highlight various features that can contribute differently to determine individual tumor classes. Some tumor classes depend more on contextual information, while in other cases, details from higher magnifications are more relevant. Predictions for each region from each magnification level are assigned with different weights that determine the influence on the final combined diagnosis. However, it is difficult to set the correct weights manually because each tumor type or even specific tumor structure can be better comprehensible from different magnification. We propose a Weighing Expert neural network that can learn to assign different weights to individual magnified Quad regions based on the present structures.

\subsubsection{Weighing Predictions Expert Design}

\begin{figure*}[htb]
    \centering
    \includegraphics[width=\textwidth]{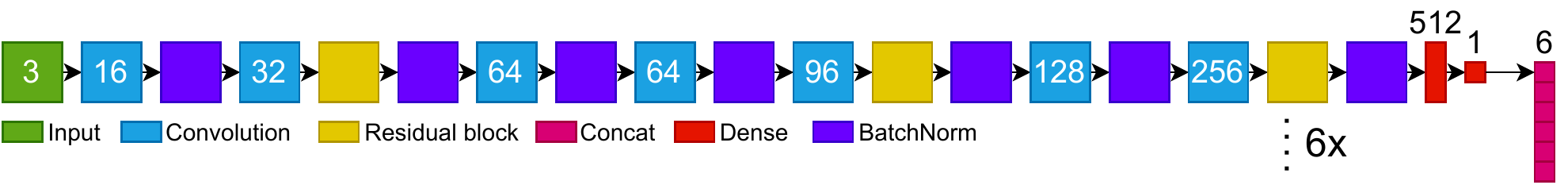}
    \caption{Simplified architecture of the Weighing Expert. The digits within Conv layers represent number of filters. Conv layers have a stride of 2, therefore the image resolution is halved with each convolution. The shown architecture until last Conv(64) is used for all six network branches, and a softmax of their summed outputs produces the final $6D$ vector of weights. Best viewed in color.}
    \label{fig:weigh_scheme}
\end{figure*}

The Expert consists of six separate networks that do not share weights, and the six products of networks are concatenated at the end to compute the final value. The simplified schema of the architecture may is in Figure \ref{fig:weigh_scheme}. Each network takes an individual input that is an image from one of six levels of T-pyramids created within the WSI Analyzer. The output is $6D$ vector - a predicted weight for each level. The training process computes the loss function from a multi-class label of ground truth annotations and six softmax labels containing the previous Expert's predictions weighted by the predicted $6D$ vector.

\begin{enumerate}
    \item 
    The loss for $N$ classes (3 classes in our work) is computed in 5 steps: 
    \item 
    The weights vector is extended to the shape of $6 \times N$. 
    \item 
    The weights vector element-wise multiplies the previous Expert's predictions producing $6 \times N$ matrix of weighted predictions. $$W_{pred} = weights \odot predictions$$
    \item 
    The summation of the weighted predictions produces $ND$ vector representing weighted multi-class label - $W_{sum}$, with each class represented in this vector having a value in range $\langle 0, \infty) $.
    \item 
    Softmax to these labels produces $W_{Softmax}$
    \item 
    Finally, the loss is computed as a categorical cross-entropy between the $gt$ labels and $W_{Softmax}$ values. 
    $$Loss = -  \sum_{i=1}^{N}gt_i . log(W_{Softmax_i})$$
\end{enumerate}

The optimizer was Adam with the learning rate of $1e^{-4}$. The mini-batch size was 128; the model trained for 55 epochs before early stopped; the best loss achieved was 0.5814, and the best evaluation loss of 1.154. The classification evaluation metrics are later in the results section.

\section{Results}

In this section, we evaluate the trained Experts. First, we evaluate individual Experts' performance, focusing purely on the Expert's designated task. In the second part, we evaluate classification on the whole \gls{WSI} exploiting the context information. All used evaluation metrics of different Experts are shown in Table \ref{tab:expert_level}.
All training runs were performed on an 6GB GPU Nvidia GTX 1060 Max-Q design.

\begin{table}[htb]
\centering
\caption{The best results achieved by our Experts. Strategy is the strategy used for evaluation: Balanced = subset of balanced classes; Slide = the selected regions of one whole slide; Flip is the measure of the flip operation.}
\label{tab:expert_level}
\resizebox{\columnwidth}{!}{%
\begin{tabular}{|c|c|c|c|c|c|c|}
\hline
Expert & Classes & Strategy & Accuracy & Precision & Recall & Specificity \\ \hline
Binary & 2 & Balanced & 88.6\% & 88.2\% & 88\% & 89.2\% \\ \hline
\multirow{2}{*}{Multi-Class} & \multirow{2}{*}{3} & Balanced & 72.2\% & 58.5\% & 58.3\% & 79.1\% \\ \cline{3-7} 
 &  & Slide & 85.3\% & 49.5\% & 64.3\% & 82.4\% \\ \hline
\multirow{2}{*}{Weighing} & \multirow{2}{*}{3} & Flip & 84.8\% & 77.5\% & 43.4\% & 96.4\% \\ \cline{3-7} 
 &  & Slide & 89.8\% & 55.1\% & 65.5\% & 84.5\% \\ \hline
\end{tabular}%
}
\end{table}

\subsection{Expert Level Results}

First, we present the results of the network that represents each Expert. We evaluate the Expert to assess how well the Experts handle the designated tasks regardless of the context information. All of the results, compared to other state-of-the-art approaches, are shown in Table \ref{tab:tab-comp}.

\begin{table}[htb]
\centering
\caption{The best achieved results of analyzed methods for classification. Last row are the best results achieved using our training strategy with state-of-the-art nets.}
\label{tab:tab-comp}
\resizebox{\columnwidth}{!}{%
\begin{tabular}{|c|c|c|c|}
\hline
Authors & Approach & \begin{tabular}[c]{@{}c@{}}Number \\of classes\end{tabular} & Accuracy \\ \hline
Spanhol et al.\cite{Spanholdataset} & \begin{tabular}[c]{@{}c@{}}PFTAS descriptor +\\ Traditional classifiers\end{tabular} & 2 & 83.33\% \\ \hline
Spanhol et al.\cite{Spanholcnn} & DCNN & 2 & 87.28\% \\ \hline
\multirow{2}{*}{Bayramoglu et al.\cite{Bayramoglu2016}} & DCNN, magnification independent & 2 & 82.1\% \\ \cline{2-4} 
 & DCNN, magnification specific & 2 & 80.66\% \\ \hline
\multirow{2}{*}{Han et al.\cite{Han2017}} & CSDCNN & 2 & 96.25\% \\ \cline{2-4} 
 & CSDCNN + Aug & 8 & 93,88\% \\ \hline
 Tripathi et al. \cite{Tripathi2021} & BiLSTM & 3 & 84.02\% \\ \hline
 \multirow{3}{*}{Our Solution} & InceptionV3 & 2 & 88.3\% \\ \cline{2-4} 
 & InceptionResnetV2 & 3 & 85.3\% \\  \cline{2-4}
 & InceptionResnetV2 + WeighingNet & 3 & 89.8\% \\ \hline 
\end{tabular}%
}
\end{table}

\subsubsection{Binary Classification Expert Results} \label{bin_exp}



The Binary Expert achieved an accuracy of 88.3\%, a precision of 91\%, and a recall of 86\% when evaluated as means of five test runs, each using 20480 previously unseen images balanced across both classes. Judging from these metrics, we can see that the Binary Expert quite successfully distinguishes between healthy and damaged tissue. These results are satisfactory enough for the Expert to filter out redundant and diagnostically unnecessary structures and mark regions required for further processing.

\subsubsection{Multi-class Classification Expert Results} \label{multi_exp}

\begin{figure}[htb]
    \centering
    \includegraphics[width=0.7\columnwidth]{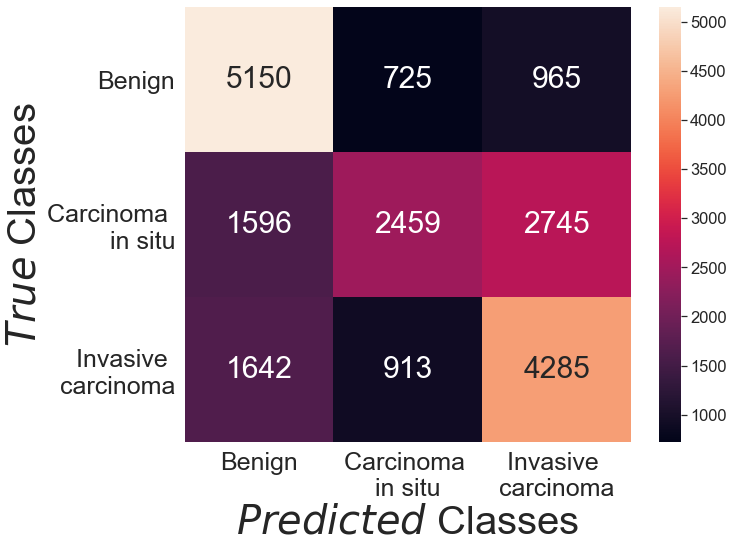}
    \caption{Confusion matrix for the Multi-Class Expert. Best viewed in color.}
    \label{fig:cm_mult}
\end{figure}

The multi-class Expert achieved an accuracy of 62.5\%, with 63\% precision and 62\% recall. The test was performed the same way with five test runs, each 20480 previously unused images, evenly distributed among all classes. Figure \ref{fig:cm_mult} shows the confusion matrix for three different predicted classes. The results show room for improvement, but as previously stated, the difficult nature of breast histopathology visuals limits the performance of any deep learning model. Contrary to Han \cite{Han2017}, who uses special architecture and loss function, we try to deal with this limit using a Weighing Expert.

\subsubsection{Weighing Predictions Expert Results} \label{weigh_exp}

In order to evaluate whether the Weighing Expert is working correctly, we define the action of changing one class to another as  Flip and the action of not changing the class as No Flip. Using the non-weighted predictions, the weighted predictions, and the ground truth annotations, we may define four scenarios:
\begin{itemize}
    \item Correct Flip - the predicted class changes from an incorrect one to a correct one (equivalent to true positive)
    \item Incorrect Flips - the predicted class changes from a correct one to an incorrect one (false positive)
    \item Correct No Flips - the predicted class does not change and stays correct (true negative)
    \item Incorrect No Flips - the predicted class does not change and stays incorrect, or the predicted class changes, but from an incorrect one to an incorrect one (false negative). 
\end{itemize}
Therefore, we can compute all usual metrics using true positives, true negatives, false positives, and false negatives. 

We performed the evaluation using 12123 samples from previously unseen \gls{WSI} where the metric measured the success rate of correctly correcting the previous Expert's outputs, in other words, if the class probability was Flipped when it was necessary. The Weighting Prediction Expert scored an accuracy of 76\%, a precision of 77\%, and a recall of 36\%. 

Weighting Prediction Expert achieved an accuracy of 75.8\%; a precision of 53\%; and a recall of 74\% in the measurement of the classification performance comparing weighted predictions to the ground truth labels.

\begin{figure}[htb]
    \centering
    \includegraphics[width=0.55\columnwidth]{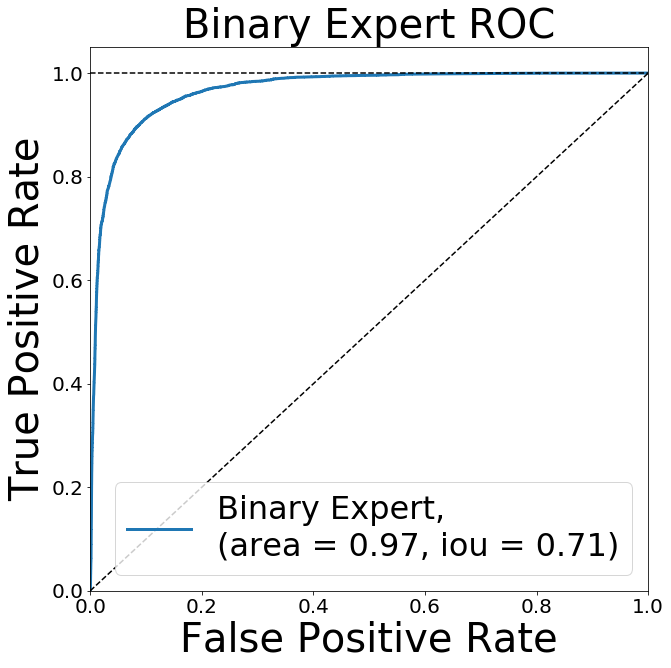}
    \caption{ROC of the Binary Expert achieved on the Evaluation slide. The dotted line = random guessing. Best viewed in color.}
    \label{fig:roc_comp}
\end{figure}

\subsection{WSI Level Results}

Now we evaluate the performance of our framework combining the results of each Expert on the classification of all regions of one \gls{WSI}. The optimal quantitative metric for this evaluation is \gls{IoU}, computed as $$IoU = \frac{gt \cap predicted}{gt \cup predicted}$$
We also empirically evaluate each analysis result from a histopathologist's perspective, using the analysis as assistance.

\subsubsection{Binary Expert WSI Level Results}

\begin{figure}[htb]
    \includegraphics[width=\columnwidth]{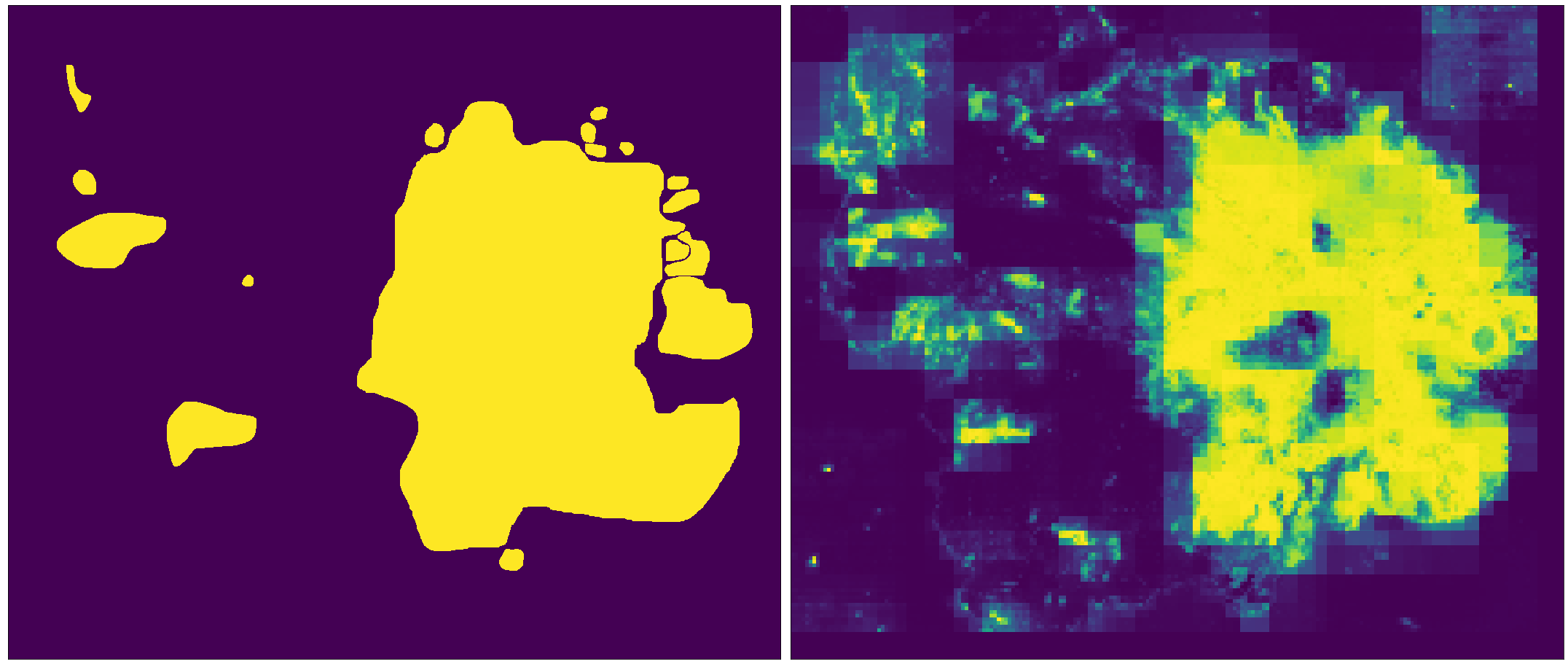}
    \caption{Left: GT binary annotations of the Slide. Right: Analysis performed by the Binary Expert. Higher intensity of the color means higher confidence of the Expert. Best viewed in color.}
    \label{fig:eval_binary}
\end{figure}

Figure \ref{fig:roc_comp} shows the ROC of the Binary Expert. 
Based on several achieved scores: 0.98 AUC, 0.74 \gls{IoU}, empirical evaluation using GT annotation, and the results of Binary Expert analysis shown in Figure \ref{fig:eval_binary}, we can establish that the Expert can distinguish between healthy and damaged tissue and filter stroma regions without significant structures. The identified possibly damaged stroma regions can be proposed to other Experts for more complex analysis.

\subsubsection{Multi-Class Expert WSI Level Results}

\begin{figure}[htb]
    \centering
    \includegraphics[width=\columnwidth]{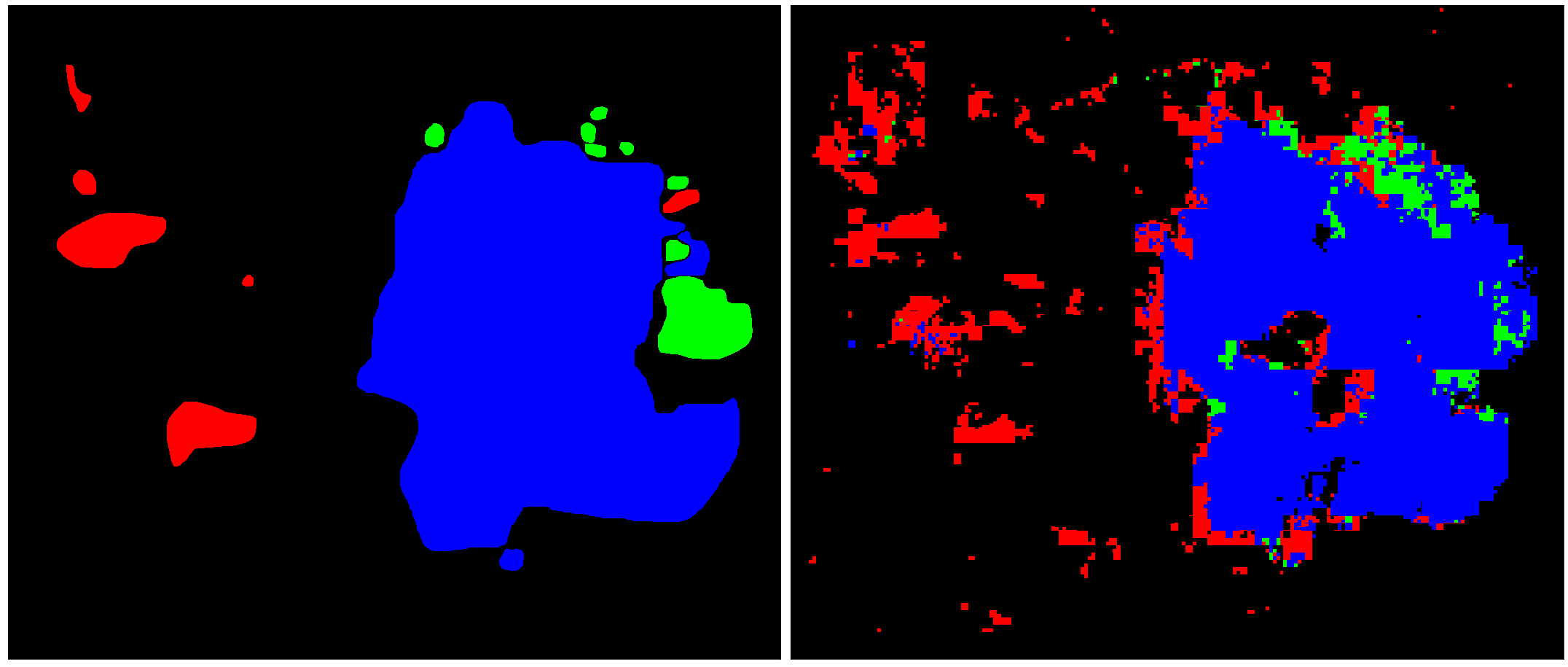}
    \caption{Left: The GT annotations. Red channel represents Benign tumor, Green channel is In Situ tumor and Blue channel is region with Invasive tumor. Right: Analysis produced by the Multi-Class Expert. Best viewed in color.} \label{fig:mult_res}
\end{figure}

Figure \ref{fig:mult_res} shows the analysis results by the Multi-class Expert separately with the ground truth (GT) annotations of searched tumor classes. \gls{IoU} scores (micro and macro average) of the whole slide are shown in Table \ref{tab:iou}. The "Multi-Class" strategy processed only GT tumor regions, hence the high classification/segmentation score. The "Multi-Class Pipeline" strategy used "Binary Expert" first and then analyzed regions marked as damaged with confidence greater than 0.3. The evaluation proved that the Binary Expert region proposals are precise enough to help make the area analyzed by subsequent Experts smaller, thus saving time and resources. Furthermore, we can see that the Expert can pick up the differences between tumor classes and create a complex analysis. The last strategy, called "Weighing Pipeline", used our weighing network to determine the contribution of individual magnifications to the final classification.

\begin{table*}[htb]
\centering
\caption{\gls{IoU} scores achieved by the Multi-Class Expert, the Multi-Class Expert Pipeline and the Weighted Pipeline when evaluated using a previously unseen \gls{WSI}.}
\label{tab:iou}
\begin{tabular}{|c|c|c|c|c|}
\hline
Class & Strategy & Micro Average & Macro Average & Binary \\ \hline
\multirow{3}{*}{All} & Multi-Class & 0.93 & 0.55 & - \\ \cline{2-5} 
 & Multi-Class Pipeline & 0.9 & 0.47 & - \\ \cline{2-5} 
 & Weighing Pipeline & 0.9 & 0.47 & - \\ \hline
\multirow{3}{*}{Benign} & Multi-Class & 0.94 & 0.66 & 0.35 \\ \cline{2-5} 
 & Multi-Class Pipeline & 0.88 & 0.54 & 0.14 \\ \cline{2-5} 
 & Weighing Pipeline & 0.88 & 0.54 & 0.14 \\ \hline
\multirow{3}{*}{In Situ} & Multi-Class & 0.95 & 0.54 & 0.11 \\ \cline{2-5} 
 & Multi-Class Pipeline & 0.95 & 0.54 & 0.1 \\ \cline{2-5} 
 & Weighing Pipeline & 0.96 & 0.53 & 0.07 \\ \hline
\multirow{3}{*}{Invasive} & Multi-Class & 0.9 & 0.85 & 0.77 \\ \cline{2-5} 
 & Multi-Class Pipeline & 0.86 & 0.8 & 0.7 \\ \cline{2-5} 
 & Weighing Pipeline & 0.87 & 0.82 & 0.73 \\ \hline
\end{tabular}%
\end{table*}

Similarly to the results of Multi-Class Expert evaluation (Section \ref{multi_exp}), we see that even though the Expert is often able to analyze regions correctly, confusion of the classes caused by all of the mentioned problems is still present, even on \gls{WSI} level.

\subsubsection{Weighing Expert WSI Level Results}

\begin{figure}[htb]
    \centering
    \includegraphics[width=\columnwidth]{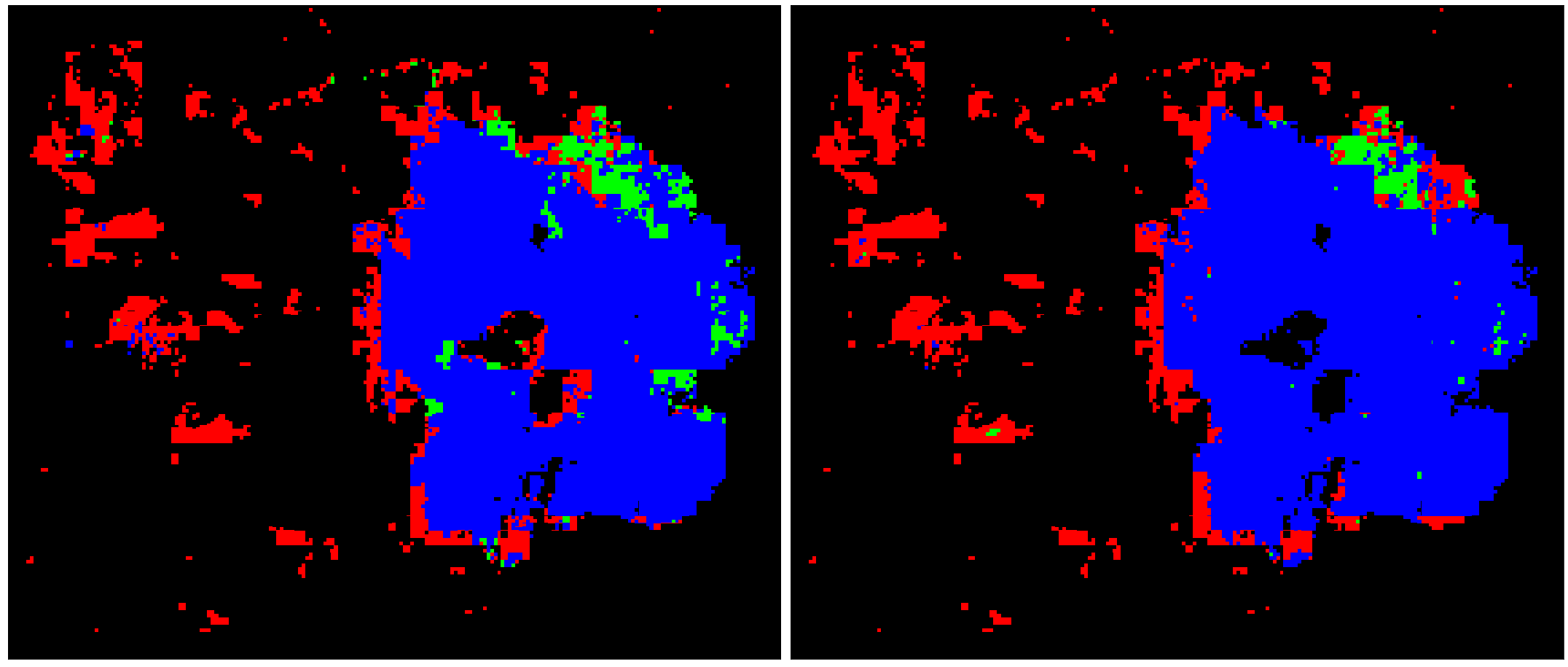}
    \caption{Analysis predictions produced by the Multi-Class Expert (left) and the same analysis predictions weighted by the Weighing Expert (right). Red channel represents Benign tumor, Green channel is In Situ tumor and Blue channel is region with Invasive tumor. Best viewed in color.} \label{fig:weighted_non_weighted}
\end{figure}

Figure \ref{fig:weighted_non_weighted} shows the comparison of the analysis produced by the Multi-Class Expert and the same analysis weighted by the Weighing Expert. The improvement of the \gls{IoU} is visible in Table \ref{tab:iou} when comparing the simple Multi-Class versus Weighing Pipeline strategy. The Weighing Expert network improves the classifications and the metric - \gls{IoU}.  

The empirical evaluation of Weighing Expert's performance by comparing the weighted analysis (Figure \ref{fig:weighted_non_weighted}, right) to non-weighted (Figure \ref{fig:weighted_non_weighted}, left) and ground truth annotations (Figure \ref{fig:mult_res}, right) shows improvements in predictions. The Weighing Expert corrects classes in the areas with wrong predictions, specifically: in the middle, a large area of the In Situ (green) region, and Benign (red) changed correctly to Invasive carcinoma (blue) class region; we also detected some minor changes in the top-left areas after closer inspection. These results suggest that the Expert is working properly, but the automatic evaluation metrics tell us that there is still room for improvement.

\section{Discussion}
As already mentioned, there are two types of publicly available datasets: the WSI images annotated by histologists (ICIAR \cite{ARESTA2019122}); and selected samples from one magnification as a chosen trade-off between context and detail \cite{Spanholdataset,pcam_dataset,pcam_dataset2}. During our work, we noted problems with both dataset types. 

The pre-selected magnification with generated samples suffers from information loss. It lacks information about the architecture of the analyzed structure. We showed the digital cut-out samples to several doctors individually, and they all agreed they could not state the diagnosis because they did not see where in the whole slice the structure was and what its surroundings were concerning the tissue morphology. 

With this knowledge, the second problem emerges. Even when its surroundings are visible, it is vague to mark the border of the tumor precisely. Almost all of the existing annotations are only humanly possible approximations with Bayesian error. With the help of histopathologists, we performed our own annotations to mark borders more precisely than the original ICIAR 2018 annotations.

\begin{figure*}[htb]
    \centering
    \includegraphics[width=\textwidth]{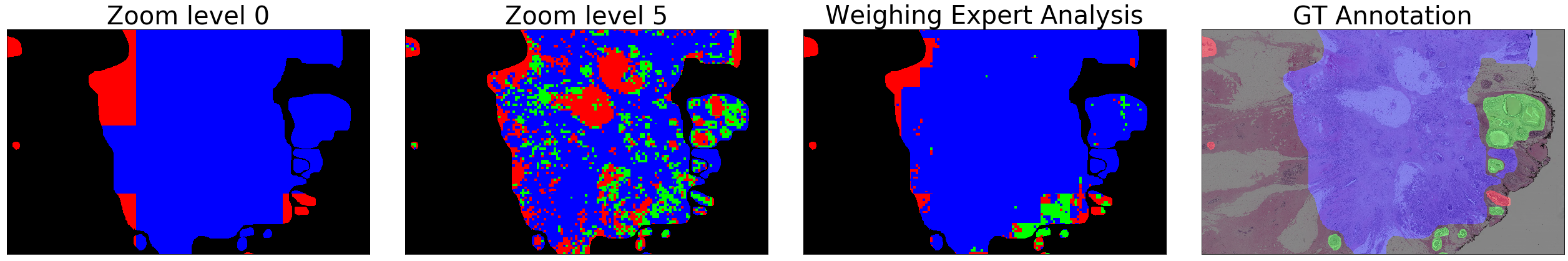}
    \caption{Images from left to right: classification on magnification level 0 (lowest); on magnification level 5 (highest); unification and reclassification by weighing Expert; ground truth annotation over tissue slice. The classification on the highest detail level inserts a high class-noise, which is later partially corrected by Weighing Expert. Unfortunately, our Weighing Expert reclassified In Situ carcinoma to Invasive as seen in the top right corner, where high detail classifications were correct, even though correctly unified region in the bottom right corner. Best viewed in color.}
    \label{fig:magnif_comp}
\end{figure*}

When the network analyzes a tiny detailed region with its architectural context, it can mark the regions precisely in even more detail than human experts, even if it is not necessary for the final diagnosis. Human experts often enclose large regions containing other tumor types in the dataset annotations because it is less manual work to select only the largest tumor region. Therefore, we do not mean that the network overperforms the histologist, but that the network only marks more small regions because it operates on the window ($256\times 256$ pixels on all magnifications) size level. In Figure \ref{fig:magnif_comp}, we can recognize predictions of benign tumor areas on the left border of the invasive region. Those errors are caused mainly through approximated annotations and similarity of stroma in benign tumor regions and borders of invasive carcinoma. On the other hand, when it processes samples from high magnification with much detail, sometimes the detail can overweight the context. Then sparse regions of different classes can appear within a large block, which we can observe in Zoom level 5 - many in situ and large benign regions classified in invasive tissue. The ground truth annotations show that benign regions are regions with fat or ducts, and in situ carcinoma is predicted without high-level context - tissue architecture.

Our weighing expert should correct these misinterpretations and unify the region if it contains only a particular tumor type, as can be seen in Figure \ref{fig:magnif_comp}. Unifying the annotation regions and the annotation should more resemble the work of a histopathologist as is also stated in \cite{Ni2019}. It is visible in Table \ref{tab:iou}, in Confusion Matrix \ref{fig:cm_mult} and in Figure \ref{fig:magnif_comp} that our multi-class expert and Weighing Expert are leaning towards the Invasive carcinoma at the expense of In Situ carcinoma. The In Situ is the only class that the results have worsen in binary classification (one class against rest). We consider this behavior a lesser problem because invasive carcinoma is much more dangerous and more deadly \cite{Ni2019}, which also relates to the unification of large regions.

In comparing our multi-scale adaptive weighted voting approach to the majority of related work analyzing only microscopy patches of single magnification, our method is more general, and it better utilizes a large scale of \gls{WSI} data. Furthermore, our selected patches of size $256\times 256$ are a standard size but contain more information than $50\times 50$ patches used in Celik et al. \cite{Celik2020} and helped to avoid overfitting. 

The usual strategy of combining the evaluated patches is majority voting \cite{Feng2021}, or thresholded or highest confidence \cite{Gandomkar2018}, which Feng et al. \cite{Feng2021} identified as problematic. Instead, our weighing network learns to change the contribution power of each magnification adaptively and improves the overall performance of the multi-class classifier pipeline from 72\% to 84\%.

\section{Conclusion}

In this paper, we present a universal framework for complex \gls{WSI} analysis, using various approaches based on deep learning models in the form of Experts. We also present multiple Experts: the Binary Expert for distinguishing between healthy and damaged parts of the \gls{WSI} and proposing damaged regions to other Experts for further analysis; the Multi-Class Classification Expert for classifying regions as different carcinoma classes and the Weighing Expert for better utilization of multiple magnifications in \gls{WSI} file. As a result of using the combined Experts, we obtain a complex and unbiased analysis of the whole \gls{WSI} that can serve as a reference or a helping tool for pathologists. 

Our binary classification approach using our custom data preparation and generation and custom trained well-known model of InceptionV3 architecture are comparable to other state-of-the-art approaches presented by Spanhol \cite{Spanholcnn}, \cite{Spanholdataset}, and Bayramoglu \cite{Bayramoglu2016}, Tripathi \cite{Tripathi2021}, and Feng \cite{Feng2021}.

Our multi-class classification results are not yet on par with the research of Han \cite{Han2017} and other related works. However, we propose an approach easy to interpret for histopathologists and similar to their workflow. Furthermore, increasing the number of trainable data, better hyper-parameter optimization of the InceptionResnetV2, further exploration of the Weighing Expert's capabilities, and increasing the number of parameters for each network the Expert contains to enhance the learning capacity may help to achieve state-of-the-art results in this domain as well in the future.

The universality of our designed framework allows the creation of an arbitrary number of Experts, each analyzing a specific part of the WSI, leading to even more complex and precise analyses. For example, an Expert for segmentation of the tissue, distinguishing even better between the classes, especially on the tumor border regions where different cancer types overlap, would be valuable, although hard to train due to the visual structure of histopathology images. Another Expert possibility would be a counter of mitoses on the WSI level, which is often required to grade the cancer stage.

\bibliographystyle{IEEEtran}
\bibliography{report}

\begin{thebibliography}{10}
\providecommand{\url}[1]{#1}
\csname url@samestyle\endcsname
\providecommand{\newblock}{\relax}
\providecommand{\bibinfo}[2]{#2}
\providecommand{\BIBentrySTDinterwordspacing}{\spaceskip=0pt\relax}
\providecommand{\BIBentryALTinterwordstretchfactor}{4}
\providecommand{\BIBentryALTinterwordspacing}{\spaceskip=\fontdimen2\font plus
\BIBentryALTinterwordstretchfactor\fontdimen3\font minus
  \fontdimen4\font\relax}
\providecommand{\BIBforeignlanguage}[2]{{%
\expandafter\ifx\csname l@#1\endcsname\relax
\typeout{** WARNING: IEEEtran.bst: No hyphenation pattern has been}%
\typeout{** loaded for the language `#1'. Using the pattern for}%
\typeout{** the default language instead.}%
\else
\language=\csname l@#1\endcsname
\fi
#2}}
\providecommand{\BIBdecl}{\relax}
\BIBdecl

\bibitem{whobreastcancer}
\BIBentryALTinterwordspacing
``Breast cancer,'' Mar 2021. [Online]. Available:
  \url{https://www.who.int/news-room/fact-sheets/detail/breast-cancer}
\BIBentrySTDinterwordspacing

\bibitem{Litjens2017}
G.~Litjens, T.~Kooi, B.~E. Bejnordi, A.~A.~A. Setio, F.~Ciompi, M.~Ghafoorian,
  J.~A. van~der Laak, B.~van Ginneken, and C.~I. S{\'{a}}nchez, ``{A survey on
  deep learning in medical image analysis},'' \emph{Medical Image Analysis},
  vol.~42, no. December 2012, pp. 60--88, 2017.

\bibitem{Spanholdataset}
F.~A. Spanhol, P.~R. Cavalin, L.~S. Oliveira, C.~Petitjean, and L.~Heutte, ``{A
  Dataset for Breast Cancer Histopathological Image Classificatio},''
  \emph{IEEE Transactions on Biomedical Engineering}, vol.~63, no.~7, pp.
  1455--1462, 2016.

\bibitem{Bayramoglu2016}
N.~Bayramoglu, ``{Deep Learning for Magnification Independent Breast Cancer
  Histopathology Image Classification},'' pp. 2440--2445, 2016.

\bibitem{Attallah2021}
O.~Attallah, F.~Anwar, N.~M. Ghanem, and M.~A. Ismail, ``{Histo-CADx: duo
  cascaded fusion stages for breast cancer diagnosis from histopathological
  images},'' \emph{PeerJ Computer Science}, vol.~7, p. e493, 2021.

\bibitem{Celik2020}
\BIBentryALTinterwordspacing
Y.~Celik, M.~Talo, O.~Yildirim, M.~Karabatak, and U.~R. Acharya, ``{Automated
  invasive ductal carcinoma detection based using deep transfer learning with
  whole-slide images},'' \emph{Pattern Recognition Letters}, vol. 133, pp.
  232--239, 2020. [Online]. Available:
  \url{https://doi.org/10.1016/j.patrec.2020.03.011}
\BIBentrySTDinterwordspacing

\bibitem{Gandomkar2018}
\BIBentryALTinterwordspacing
Z.~Gandomkar, P.~C. Brennan, and C.~Mello-Thoms, ``{MuDeRN: Multi-category
  classification of breast histopathological image using deep residual
  networks},'' \emph{Artificial Intelligence in Medicine}, vol.~88, pp. 14--24,
  2018. [Online]. Available: \url{https://doi.org/10.1016/j.artmed.2018.04.005}
\BIBentrySTDinterwordspacing

\bibitem{Togacar2020}
\BIBentryALTinterwordspacing
M.~Toğa{\c{c}}ar, K.~B. {\"{O}}zkurt, B.~Ergen, and Z.~C{\"{o}}mert,
  ``{BreastNet: A novel convolutional neural network model through
  histopathological images for the diagnosis of breast cancer},'' \emph{Physica
  A: Statistical Mechanics and its Applications}, vol. 545, p. 123592, 2020.
  [Online]. Available: \url{https://doi.org/10.1016/j.physa.2019.123592}
\BIBentrySTDinterwordspacing

\bibitem{Han2017}
\BIBentryALTinterwordspacing
Z.~Han, B.~Wei, Y.~Zheng, Y.~Yin, K.~Li, and S.~Li, ``{Breast Cancer
  Multi-classification from Histopathological Images with Structured Deep
  Learning Model},'' \emph{Scientific Reports}, vol.~7, no.~1, p. 4172, 2017.
  [Online]. Available: \url{https://doi.org/10.1038/s41598-017-04075-z}
\BIBentrySTDinterwordspacing

\bibitem{Tripathi2021}
S.~Tripathi, S.~K. Singh, and H.~K. Lee, ``{An end-to-end breast tumour
  classification model using context-based patch modelling – A BiLSTM
  approach for image classification},'' \emph{Computerized Medical Imaging and
  Graphics}, vol.~87, p. 101838, jan 2021.

\bibitem{Gecer2018}
\BIBentryALTinterwordspacing
B.~Gecer, S.~Aksoy, E.~Mercan, L.~G. Shapiro, D.~L. Weaver, and J.~G. Elmore,
  ``{Detection and classification of cancer in whole slide breast
  histopathology images using deep convolutional networks},'' \emph{Pattern
  Recognition}, vol.~84, pp. 345--356, dec 2018. [Online]. Available:
  \url{https://www.sciencedirect.com/science/article/pii/S0031320318302577}
\BIBentrySTDinterwordspacing

\bibitem{Ni2019}
\BIBentryALTinterwordspacing
H.~Ni, H.~Liu, K.~Wang, X.~Wang, X.~Zhou, and Y.~Qian, ``{WSI-Net: Branch-Based
  and Hierarchy-Aware Network for Segmentation and Classification of Breast
  Histopathological Whole-Slide Images},'' in \emph{Lecture Notes in Computer
  Science (including subseries Lecture Notes in Artificial Intelligence and
  Lecture Notes in Bioinformatics)}, vol. 11861 LNCS.\hskip 1em plus 0.5em
  minus 0.4em\relax Springer, oct 2019, pp. 36--44. [Online]. Available:
  \url{https://doi.org/10.1007/978-3-030-32692-0\_5}
\BIBentrySTDinterwordspacing

\bibitem{Feng2021}
Y.~Feng, A.~Hafiane, and H.~Laurent, ``A deep learning based multiscale
  approach to segment the areas of interest in whole slide images,''
  \emph{Computerized Medical Imaging and Graphics}, vol.~90, p. 101923, 2021.

\bibitem{ARESTA2019122}
\BIBentryALTinterwordspacing
G.~Aresta, T.~Araújo, S.~Kwok, S.~S. Chennamsetty, M.~Safwan, V.~Alex,
  B.~Marami, M.~Prastawa, M.~Chan, M.~Donovan, G.~Fernandez, J.~Zeineh,
  M.~Kohl, C.~Walz, F.~Ludwig, S.~Braunewell, M.~Baust, Q.~D. Vu, M.~N.~N. To,
  E.~Kim, J.~T. Kwak, S.~Galal, V.~Sanchez-Freire, N.~Brancati, M.~Frucci,
  D.~Riccio, Y.~Wang, L.~Sun, K.~Ma, J.~Fang, I.~Kone, L.~Boulmane,
  A.~Campilho, C.~Eloy, A.~Polónia, and P.~Aguiar, ``Bach: Grand challenge on
  breast cancer histology images,'' \emph{Medical Image Analysis}, vol.~56, pp.
  122--139, 2019. [Online]. Available:
  \url{https://www.sciencedirect.com/science/article/pii/S1361841518307941}
\BIBentrySTDinterwordspacing

\bibitem{Li2019}
W.~Li, J.~Li, K.~V. Sarma, K.~C. Ho, S.~Shen, B.~S. Knudsen, A.~Gertych, and
  C.~W. Arnold, ``{Path R-CNN for Prostate Cancer Diagnosis and Gleason Grading
  of Histological Images},'' \emph{IEEE Transactions on Medical Imaging},
  vol.~38, no.~4, pp. 945--954, 2019.

\bibitem{szegedy2016rethinking}
C.~Szegedy, V.~Vanhoucke, S.~Ioffe, J.~Shlens, and Z.~Wojna, ``Rethinking the
  inception architecture for computer vision,'' in \emph{Proceedings of the
  IEEE conference on computer vision and pattern recognition}, 2016, pp.
  2818--2826.

\bibitem{szegedy2017inception}
C.~Szegedy, S.~Ioffe, V.~Vanhoucke, and A.~Alemi, ``Inception-v4,
  inception-resnet and the impact of residual connections on learning,'' in
  \emph{Proceedings of the AAAI Conference on Artificial Intelligence},
  vol.~31, no.~1, 2017.

\bibitem{Spanholcnn}
\BIBentryALTinterwordspacing
F.~A. Spanhol, L.~S. Oliveira, C.~Petitjean, and L.~Heutte, ``{Breast cancer
  histopathological image classification using Convolutional Neural
  Networks},'' in \emph{2016 International Joint Conference on Neural Networks
  (IJCNN)}.\hskip 1em plus 0.5em minus 0.4em\relax IEEE, 07 2016, pp.
  2560--2567. [Online]. Available:
  \url{http://ieeexplore.ieee.org/document/7727519/}
\BIBentrySTDinterwordspacing

\bibitem{pcam_dataset}
\BIBentryALTinterwordspacing
B.~Ehteshami~Bejnordi, M.~Veta, P.~Johannes~van Diest, B.~van Ginneken,
  N.~Karssemeijer, G.~Litjens, J.~A. W.~M. van~der Laak, , and the
  CAMELYON16~Consortium, ``{Diagnostic Assessment of Deep Learning Algorithms
  for Detection of Lymph Node Metastases in Women With Breast Cancer},''
  \emph{JAMA}, vol. 318, no.~22, pp. 2199--2210, 12 2017. [Online]. Available:
  \url{https://doi.org/10.1001/jama.2017.14585}
\BIBentrySTDinterwordspacing

\bibitem{pcam_dataset2}
B.~S. Veeling, J.~Linmans, J.~Winkens, T.~Cohen, and M.~Welling, ``Rotation
  equivariant cnns for digital pathology,'' in \emph{International Conference
  on Medical image computing and computer-assisted intervention}.\hskip 1em
  plus 0.5em minus 0.4em\relax Springer, 2018, pp. 210--218.

\end{thebibliography}

\end{document}